\documentclass[pra, twocolumn]{revtex4-1}

\usepackage{amsmath, amssymb, bm}
\usepackage{graphicx}
\usepackage{hyperref}
\usepackage{xcolor}
\usepackage[english]{babel}

\begin{document}
\title{Noise-resilient Universal Quantum Computing in the Presence of Anisotropic Noise}

\author{Yang-Yang Xie$^{1, 4}$}

\author{Zhao-Ming Wang$^{2, 3, 4}$}
\email{wangzhaoming@ouc.edu.cn}

\author{Lian-Ao Wu$^{1, 5, 6}$}
\email{lianaowu@gmail.com}

\affiliation{
$^{1}$ Department of Physics, University of the Basque Country UPV/EHU, Bilbao 48080, Spain \\
$^{2}$ Engineering Research Center of Advanced Marine Physical Instruments and Equipment of Ministry of Education, Ocean University of China, Qingdao 266100, China \\
$^{3}$ Qingdao Key Laboratory of Optics and Optoelectronics, Qingdao 266100, China \\
$^{4}$ College of Physics and Optoelectronic Engineering, Ocean University of China, Qingdao 266100, China \\
$^{5}$ IKERBASQUE Basque Foundation for Science, Bilbao 48013, Spain \\
$^{6}$ EHU Quantum Center, University of the Basque Country UPV/EHU, Bilbao 48940, Spain
}

\date{\today}

\begin{abstract}
We propose a universal gate set for quantum computing that operates in the presence of decoherence without the overhead of active error correction. We show that a broad class of anisotropic system--bath couplings can be effectively decoupled by preparing an appropriate system--bath entangled initial state. The initially established entanglement serves as a resource to cancel out the dominant decoherence during evolution, enabling quantum computation to proceed as if the system were effective decoupled from its environment.
\end{abstract}

\maketitle

\section{Introduction}
Quantum computing promises to solve problems far beyond the reach of even the most powerful classical supercomputers. In recent years, this promise has fueled rapid experimental progress across multiple platforms, including superconducting circuits~\cite{arute2019}, trapped ions~\cite{pogorelov2021}, semiconductor systems~\cite{petta2005}, and photonic systems~\cite{larsen2025}. However, their practical performance remains fundamentally constrained by decoherence, the loss of coherence from unavoidable system--environment interactions~\cite{breuer2002, pellizzari1995}.

A wide range of strategies have been developed to cope with decoherence. Among them, quantum error correction (QEC) stands out as the leading paradigm~\cite{shor1995, knill1998}. QEC safeguards logical qubits by first encoding the information into a redundant set of physical qubits, and then applying repeated rounds of active error correction. The recent demonstration of the surface code on Google's Willow processor represents a promising step forward~\cite{acharya2025}, yet it also implicitly reveals the substantial overhead: often requiring hundreds of physical qubits per logical qubit~\cite{rini2024}. Other well-established strategies provide different approaches. For example, dynamical decoupling (DD)~\cite{viola1999} applies external controls to cancel noise effects, and decoherence-free subspace (DFS)~\cite{lidar1998, lidar2003, quiroz2024} encodes information into a subspace that is naturally immune to them. While effective under specific conditions, the applicability of these strategies is constrained by their stringent requirements, such as precise timing, fine-grained control, or specific symmetries in interactions~\cite{viola1999, lidar1998, lidar2003}. In general, most strategies implicitly assume an initially factorized system--bath state and fight against decoherence through active, time-dependent control during computation.

In this work, we introduce a new framework for noise-resilient quantum computing. This proposal incorporates system--bath entanglement as a resource to cancel out the main decoherence during evolution, which parallels the philosophy of one-way quantum computing~\cite{raussendorf2001, raussendorf2003, walther2005, raussendorf2007}. Specifically, we prepare an appropriate system--bath entangled initial state~\cite{wu2003, wu2024,wufr}, and the pre-established entanglement is then consumed by successive gate operations on the system during computation. As a result, the system evolves as if decoupled from the environment, thereby performing quantum computation without additional overhead. This work presents the general framework of our proposal and its application to spin--boson models with anisotropic couplings. Such anisotropic system–bath interactions naturally exist in many solid-state platforms~\cite{rower2025, dresselhaus1955, molenkamp2001, schliemann2003}.

\section{Anisotropic Noise}
The spin--boson model~\cite{leggett1987} provides a natural framework for analyzing decoherence, as qubits serve as the fundamental units of quantum computation and bosonic modes are commonly used to model environmental degrees of freedom. Here we consider a class of spin--boson models with anisotropic couplings~\cite{xie2014}. Such models are derived from the minimal-coupling Hamiltonian $H = (\boldsymbol{p} - \boldsymbol{A})^2 /(2m) + V(|\boldsymbol{r} - \boldsymbol{R}|)$, which describes a charged particle interacting with a quantized electromagnetic field, by expanding to leading order in the dipole approximation~\cite{wu2024}. The resulting effective Hamiltonian is
\begin{equation}
	H = H_\mathrm{S} + H_\mathrm{B} - \boldsymbol{d} \cdot \boldsymbol{E} - \boldsymbol{\mu} \cdot \boldsymbol{B},
	\label{h_eff}
\end{equation}
with $H_\mathrm{S}$ and $H_\mathrm{B}$ denoting the system and bath Hamiltonians, and $\boldsymbol{d}$, $\boldsymbol{\mu}$ the system's electric and magnetic dipole operators. The quantized electric and magnetic fields are expressed as $\boldsymbol{E} = \sum_k (b_k+b_k^\dagger)$ and $\boldsymbol{B} = \sum_k i(b_k - b_k^\dagger)$, where $b_k$ ($b_k^\dagger$) is the annihilation (creation) operator of mode $k$. To analyze this model in the qubit basis, we project the dipole operators onto the eigenbasis of the qubit Hamiltonian as $\boldsymbol{d} = \sum_k d_k \sigma_x$ and $\boldsymbol{\mu} = \sum_k \mu_k \sigma_y$, where $\sigma_{x,y,z}$ denote the Pauli operators of the qubit and $d_k$, $\mu_k$ are mode-dependent coupling strengths~\cite{xie2014}. Substituting these expansions into the interaction terms yields the full Hamiltonian
\begin{align}
	H = \, & \frac{\omega_{0}}{2}\sigma_z + \sum_k\omega_kb_k^{\dagger}b_k 
	+  \sum_k d_k \sigma_x \left(b_k+b_k^{\dagger}\right) \notag \\
	& +  \sum_k i\mu_k  \sigma_y \left(b_k-b_k^{\dagger}\right),
	\label{h_multimode}
\end{align}
with $\omega_0$ denoting the qubit splitting and $\omega_k$ the frequencies of the bosonic modes. Eq.~(\ref{h_multimode}) provides a general microscopic origin for anisotropic couplings: the electric and magnetic dipole terms respectively couple the $x$- and $y$-axes of the qubit to the amplitude and phase quadratures of the field. In practice, the same structure can also model other orthogonal noise channels beyond electromagnetic fields. Such anisotropy is relevant to various experimental platforms. For example, in superconducting qubits, the flux and charge operators map to orthogonal axes on the Bloch sphere owing to their canonical conjugacy, leading to anisotropic transverse noise~\cite{rower2025}. In semiconductors with spin--orbit coupling, the intrinsic Rashba and Dresselhaus effects naturally give rise to anisotropic interactions~\cite{dresselhaus1955, molenkamp2001, schliemann2003}. These can be further tuned (via external electric fields for Rashba coupling and heterostructure geometry for Dresselhaus coupling) to more closely match the anisotropic model of Eq.~(\ref{h_multimode}).

To make the role of anisotropy explicit, we rewrite the interaction Hamiltonian in terms of the raising and lowering operators, which naturally separate the rotating and counter-rotating contributions:
\begin{align}
	H = \frac{\omega_{0}}{2}\sigma_{z} + \sum_k g_k\left[\sigma_{+}b_k+\sigma_{-}b_k^{\dagger}+\lambda_k\left(\sigma_{+}b_k^{\dagger}+\sigma_{-}b_k\right)\right],
	\label{mostimportant}
\end{align}
with the overall coupling strength $g_k=(d_k+\mu_k)/2$ and the anisotropy parameter $\lambda_k=(d_k-\mu_k)/(d_k+\mu_k)$. Experimentally, for a potential platform of our proposal, one can employ techniques developed in quantum noise spectroscopy (QNS) to obtain spectral density in the system--bath interaction~\cite{sung2021, gonzalez2022, qin2024, milne2021}, namely $g_k$ and $\lambda_k$ as functions of $\omega_k$. This anisotropic model has demonstrated direct experimental implementations of $\lambda\approx0.5$ in superconducting circuits~\cite{xie2014, forn-diaz2010}. Moreover, when $\lambda=0$, it reduces to the isotropic Jaynes--Cummings model. For other values of $\lambda$, such couplings may already exist in physical systems yet remain to be fully characterized.

\section{Dressing Transformation}
Analogous to one-way quantum computing, where qubits are pre-entangled to form a resource initial state, this proposal entangles the qubit with bosonic modes for effective decoupling. Now we consider a unitary dressing transformation to analyze the transformed dynamics, which satisfies the \textit{first-order} decoupling condition
\begin{equation}
	V \left(H_\mathrm{S} + H_\mathrm{B}\right) V^\dagger - \left(H_\mathrm{S} + H_\mathrm{B}\right) \approx H_\mathrm{SB}.
	\label{dc}
\end{equation}
This condition ensures that the transformed free Hamiltonian $V H_0 V^\dagger$ incorporates the first-order interactions as an effective renormalization of the free Hamiltonian $H_0 = H_\mathrm{S} + H_\mathrm{B}$, thereby cancelling out the dominant decoherence.

Inspired by the exactly solvable dephasing model in Ref.~\cite{breuer2002}, we propose a displacement-type transformation
\begin{equation}
	V = \exp \left[ \frac{1}{2} \sigma_y \sum_k \left( \alpha_k b_k^\dagger - \alpha_k^* b_k \right) \right],
	\label{v}
\end{equation}
with $\alpha_k$ the complex parameters controlling the $\sigma_y$ coupling to each bath mode $k$. Applying $V$ to the Hamiltonian in Eq.~(\ref{h_multimode}), we obtain
\begin{equation}
	V^\dagger H V \approx \frac{\omega_0}{2} \sigma_z + \sum_k \omega_k b_k^\dagger b_k 
	= H_\mathrm{S} + H_\mathrm{B},
	\label{trans_to_free}
\end{equation}
effectively canceling out the system--bath interaction to first order.

The transformation in Eq.~(\ref{trans_to_free}) can be derived as follows. To apply $V$ to the system Hamiltonian, we rewrite it as $ V = \exp (i \sigma_y \theta/2) $, with $ \theta = \sum_k (-i \alpha_k b_k^\dagger + i \alpha_k^* b_k)$. Assuming $|\alpha_k| \ll 1$, we expand $V H_\mathrm{S} V^\dagger$ in the small-angle limit as
$ V H_\mathrm{S} V^\dagger = \frac{\omega_0}{2} (\sigma_z \cos\theta - \sigma_x \sin\theta) \approx \frac{\omega_0}{2} (\sigma_z - \sigma_x \theta) $,
where higher-order terms in $\theta$ are neglected. Substituting the expression for $\theta$, we obtain 
$ V H_\mathrm{S} V^\dagger \approx \frac{\omega_0}{2} \sigma_z + \sum_k i \frac{\omega_0}{2} \sigma_x (\alpha_k b_k^\dagger - \alpha_k^* b_k) $. 
For the bath Hamiltonian, as each mode-specific $V_k$ acts as a displacement operator 
$ V_k = \exp ( \sigma_y \alpha_kb_k^{\dagger}/2 - \sigma_y \alpha_k^{*}b_k/2 ) = D_k ( \sigma_y \alpha_k /2 ) $, this yields 
$ V H_{\mathrm{B}} V^{\dagger} =  \sum_k\omega_kb_k^{\dagger}b_k- \sum_k \frac{\omega_k}{2} \sigma_y ( \alpha_kb_k^{\dagger}+\alpha_k^{*}b_k ) $.
Combining these contributions, we obtain the transformed free Hamiltonian
\begin{align}
	V H_0 V^\dagger 
	= & \frac{\omega_{0}}{2}\sigma_z + \sum_k \omega_k b_k^{\dagger}b_k +  \sum_k i\frac{\omega_{0}\alpha_k}{2} \sigma_x \left(b_k+b_k^{\dagger}\right) \notag \\ 
	& + \sum_k\frac{\omega_k\alpha_k}{2} \sigma_y \left(b_k-b_k^{\dagger}\right),
	\label{h_full}
\end{align}
exactly recovering the anisotropic spin--boson form of Eq.~(\ref{mostimportant}) through $g_k=(\omega_0+\omega_k)i\alpha_k/2$ and $\lambda_k=(\omega_0-\omega_k)/(\omega_0+\omega_k)$. Hence, $V \left(H_\mathrm{S} + H_\mathrm{B}\right) V^\dagger = H$. By unitarity of $V$, the inverse transformation yields $V^\dagger H V = H_\mathrm{S} + H_\mathrm{B}$, which is the decoupling condition of Eq.~(\ref{dc}).

The overall coupling strengths $g_k$ observed in QNS experiments can always be reproduced via $g_k=(\omega_0+\omega_k)i\alpha_k/2$ because the parameters $\alpha_k$ can be freely chosen. In contrast, the anisotropy  $\lambda_k$ observed in QNS experiments may not be reproducible by the specific class $(\omega_0-\omega_k)/(\omega_0+\omega_k)$. Therefore, the suitable experimental platforms for our proposal are those physical systems that naturally exhibit, or can be tuned to exhibit, the anisotropy $\lambda_k = (\omega_0-\omega_k)/(\omega_0+\omega_k)$. Encouragingly, as is typical for master-equation descriptions of open quantum systems, the system dynamics (such as quantum computing) depend only weakly on the precise functional form of the anisotropy $\lambda_k$, particularly in Markovian regimes.

\section{Universal Quantum Computing}
Thus far, we have focused on a single-qubit Hamiltonian $H_\mathrm{S} = \omega_0 \sigma_z/2$ coupled to an anisotropic bath. We now generalize this framework to a multi-qubit setting that supports a universal gate set. Specifically, we assume that single-qubit $Y$ rotations and two-qubit $YY$ interactions are available, as generally realized in experimental platforms such as trapped ions~\cite{milburn2000, leibfried2003} and superconducting circuits~\cite{kjaergaard2020}. The resulting Hamiltonian is
\begin{equation}
	H_\mathrm{S} = \sum_i \frac{\omega_i}{2} \sigma_z^i + \sum_i \eta_i(t) \sigma_y^i + \sum_{i,j} J_{i,j}(t) \sigma_y^i \sigma_y^j,
	\label{hs_closed}
\end{equation}
where $\omega_i$ denote qubit frequencies, and $\eta_i(t)$, $J_{i,j}(t)$ are tunable control parameters. Free evolution under $\sigma_z^i$ implements $Z$ rotations, and the term $\eta_i(t) \sigma_y^i$ allows arbitrary rotations around the $y$-axis. Together, these operations constitute a universal single-qubit gate set. The coupling terms $\sigma_y^i \sigma_y^j$ generate entangling interactions via controlled evolution under the tunable $J_{i,j}(t)$, as implemented through Mølmer–Sørensen couplings in trapped ions~\cite{milburn2000, leibfried2003} or flux-tunable interactions in superconducting circuits~\cite{kjaergaard2020}. In summary, the Hamiltonian in Eq.~(\ref{hs_closed}) supports universal quantum computation in the absence of decoherence. In practice, however, inevitable system--bath coupling breaks down this ideal framework.

We now show that universal quantum computation remains achievable for the anisotropic system–bath interaction in Eq.~(\ref{h_multimode}) by extending the dressing transformation. For an $N$-qubit system, the global dressing transformation takes the form
\begin{align}
	V = \prod_{i=1}^{N} \exp\left[\frac{1}{2}\sigma_y^i \sum_k \left(\alpha_{i,k} b_k^\dagger - \alpha_{i,k}^* b_k\right)\right]. \notag
\end{align}
It preserves the local structure of the single-qubit transformation while embedding multi-qubit correlations with the bosonic bath. Importantly, $V$ commutes with both $\sigma_y^i$ and $\sigma_y^i \sigma_y^j$, $[V, \sigma_y^i] = 0$, $[V, \sigma_y^i \sigma_y^j] = 0$, ensuring that these gates are invariant under the transformation, preserving their exact dynamics in the dressed basis.

We now extend the framework of Eq.~(\ref{hs_closed}) to include environmental effects through the full Hamiltonian $H = H_\mathrm{S}  + H_\mathrm{B} + H_\mathrm{SB}$, with 
\begin{align}
	H_\mathrm{SB} = \sum_{i,k} \left[ i\frac{ \omega_i \alpha_{i,k}}{2} \sigma_x^i \left(b_k + b_k^\dagger\right) + \frac{\omega_k \alpha_{i,k}}{2} \sigma_y^i \left(b_k - b_k^\dagger\right) \right]. \notag
\end{align}
Applying the global transformation $V$ yields $V^\dagger H V \approx H_\mathrm{S} + H_\mathrm{B}$ to first order in the coupling parameters $\alpha_{i,k}$. In the dressed basis, the system--bath interaction is effectively eliminated, and the system evolves as if isolated from the bath.

Our proposal begins with the preparation of the system--bath entangled initial state $|\Psi^\mathrm{d}(0)\rangle = V |\Psi^\mathrm{id}(0)\rangle$. Here the superscripts ``id" and ``d" denote states in the ideal and dressed bases, respectively. To prepare this entangled state, we start from a product state in the ideal basis $|\Psi^\text{id}(0)\rangle = |\psi_\mathrm{S}\rangle \otimes | \psi_\mathrm{B}\rangle$. As in conventional quantum computing proposals, we consider a product state in which each qubit is prepared in the computational state
\begin{align}
	|0\rangle_i= \frac{|y_+\rangle_i + |y_-\rangle_i}{\sqrt{2}}, \quad \sigma_y^i|y_\pm\rangle_i = \pm|y_\pm\rangle_i, \notag
\end{align}
and each bosonic (phononic) mode is initialized in a thermal state
\begin{align}
	|\psi_{\mathrm{B},k}\rangle = \sum_{n=0}^{\infty} \sqrt{p_{n,k}}|n\rangle_{\mathrm{B}_k} |n\rangle_{\mathrm{A}_k}, \quad p_{n,k} = \frac{e^{-\beta\omega_k n}}{Z_k}. \notag
\end{align}
Here the explicit form of the thermal state is given for completeness; the subsequent discussion does not rely on its detailed structure. Then the initialization is completed by preparing the corresponding system--bath entangled initial state, as characterized by the dressing transformation. Such an initialization ensures that
\begin{align}
	|\Psi^\mathrm{d}(t)\rangle \approx V |\Psi^\mathrm{id}(t)\rangle, \quad |\Psi^{\mathrm{id}}(t)\rangle = e^{-i\left( H_\mathrm{S} + H_\mathrm{B} \right)t} |\Psi^{\mathrm{id}}(0)\rangle, \notag
\end{align}
i.e., the dressed evolution follows the ideal noise-free dynamics within the validity of the first-order approximation.

It is noteworthy that, for each qubit, initializing in the state $(|y_+\rangle + |y_-\rangle)/\sqrt{2}$ and applying the system--bath entanglement produce the entangled state
\begin{align}
	\frac{1}{\sqrt{2}} \Big[ |y_+\rangle \prod_k D_k \left( \frac{\alpha_k}{2} \right) + |y_-\rangle \prod_k D_k \left( -\frac{\alpha_k}{2} \right) \Big] |\psi_{\mathrm{B}}\rangle, \notag
\end{align}
where $D_k$ is the standard phonon displacement operator and $\alpha_k = -(i 2 g_k)/(\omega_0+\omega_k)$, as observed in QNS experiments. The simplest initial state preparation for each qubit is to start in the state $|y_+\rangle$ and after applying $V$, the system and bath are in an initial state
\begin{align}
	|\Psi^\mathrm{d}(0)\rangle=|y_+\rangle \prod_k D_k \left( \frac{\alpha_k}{2} \right) |\psi_{\mathrm{B}}\rangle, \notag
\end{align}
or in the conventional density matrix notation, 
\begin{align}
	\rho^\mathrm{d}(0)=|y_+\rangle \langle y_+|  \prod_k D_k \left( \frac{\alpha_k}{2} \right) \rho_\mathrm{B} % \prod_k 
	D^{\dagger}_k \left( \frac{\alpha_k}{2} \right), \notag
\end{align}
where $\rho_\mathrm{B}$ is the bath thermal initial state. In this specific initial-state preparation, the system and the bath are separated and the system is in $|y_+\rangle$. Promisingly, preparing the entangled state $|\Psi^\mathrm{d}(0)\rangle$ or $\rho^\mathrm{d}(0)$ reduces to creating displaced phonon number states, a task that has already been realized in numerous experiments~\cite{ziesel2013}. Remarkably, the final readout for each qubit can be readily performed in the $  |y_\pm \rangle$ basis.  This substantially enhances the viability of our proposal.

\section{Numerical Simulation}
The preceding analysis shows that the dressing transformation effectively decouples system–bath interactions to first order. To test the validity of this first-order approximation, we numerically compare the dressed dynamics with the ideal noise-free evolution. As the transformation commutes with single-qubit $Y$ rotations and two-qubit $YY$ interactions, these gates evolve exactly in the dressed basis. Thus, it is sufficient to examine the remaining $Z$ rotations.

\begin{figure}[!tb]
	\centering
	\includegraphics[width=0.5\columnwidth]{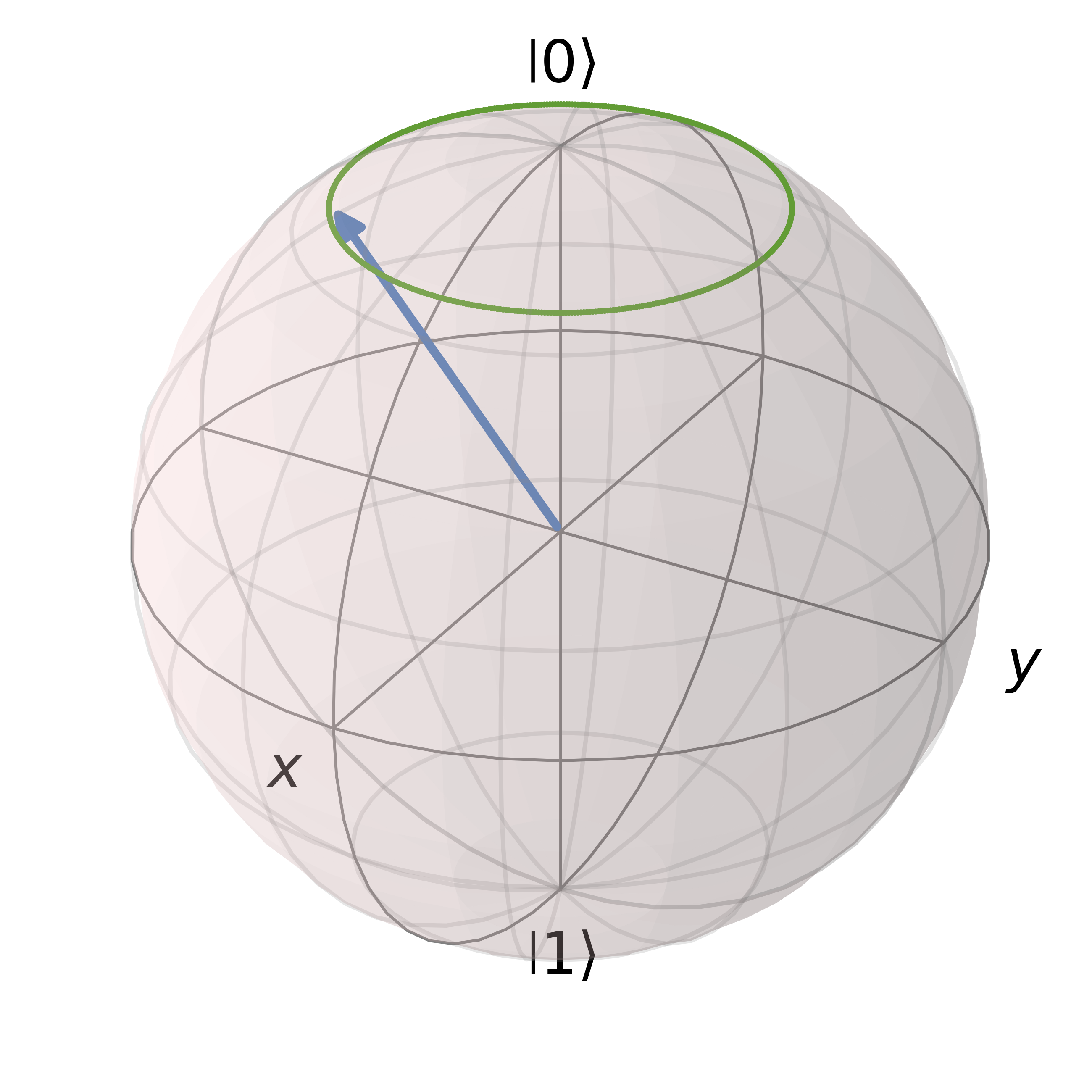}
	\caption{Representative random pure state on the Bloch sphere, used as the initial condition for fidelity evaluation. The results are qualitatively independent of the specific initial state.}
	\label{fig1}
\end{figure}
To capture the essential features of the dressed dynamics, the environment is modeled by a single bosonic mode. This reduction is justified when the spectral density is narrow or structured, such that an effective mode dominates the system--bath coupling~\cite{breuer2002, chin2010}. The combined system is initialized in the dressed basis as 
\begin{align}
	\rho(0) = V \left( |\psi_\mathrm{S}(0)\rangle\langle\psi_\mathrm{S}(0)| \otimes \rho_\mathrm{B} \right) V^\dagger, \notag
\end{align}
where $|\psi_\mathrm{S}(0)\rangle$ is a random pure state (illustrated in Fig.~\ref{fig1}), and $\rho_{\mathrm{B}} = \exp(-\beta H_{\mathrm{B}})/Z$ is a thermal state at temperature $T=1$ (with $K_{\mathrm{B}}=1$). The reduced qubit state at time $t$ is obtained as
\begin{align}
	\rho_\mathrm{S}(t) = \mathrm{Tr}_\mathrm{B} \left[ e^{-iHt} \rho(0) e^{iHt} \right], \notag
\end{align}
where $H$ is the full Hamiltonian in Eq.~(\ref{h_full}). The accuracy of the approximation is quantified by the fidelity
\begin{align}
	F(t) = \sqrt{ \left| \langle \psi_\mathrm{S}(t) | \rho_\mathrm{S}(t) | \psi_\mathrm{S}(t) \rangle \right|}, \notag
\end{align}
with $|\psi_\mathrm{S}(t)\rangle = \exp(-iH_\mathrm{S}t)|\psi_\mathrm{S}(0)\rangle$ denoting the ideal noise-free evolution.

\begin{figure}[!tb]
	\centering
	\includegraphics[width=\columnwidth]{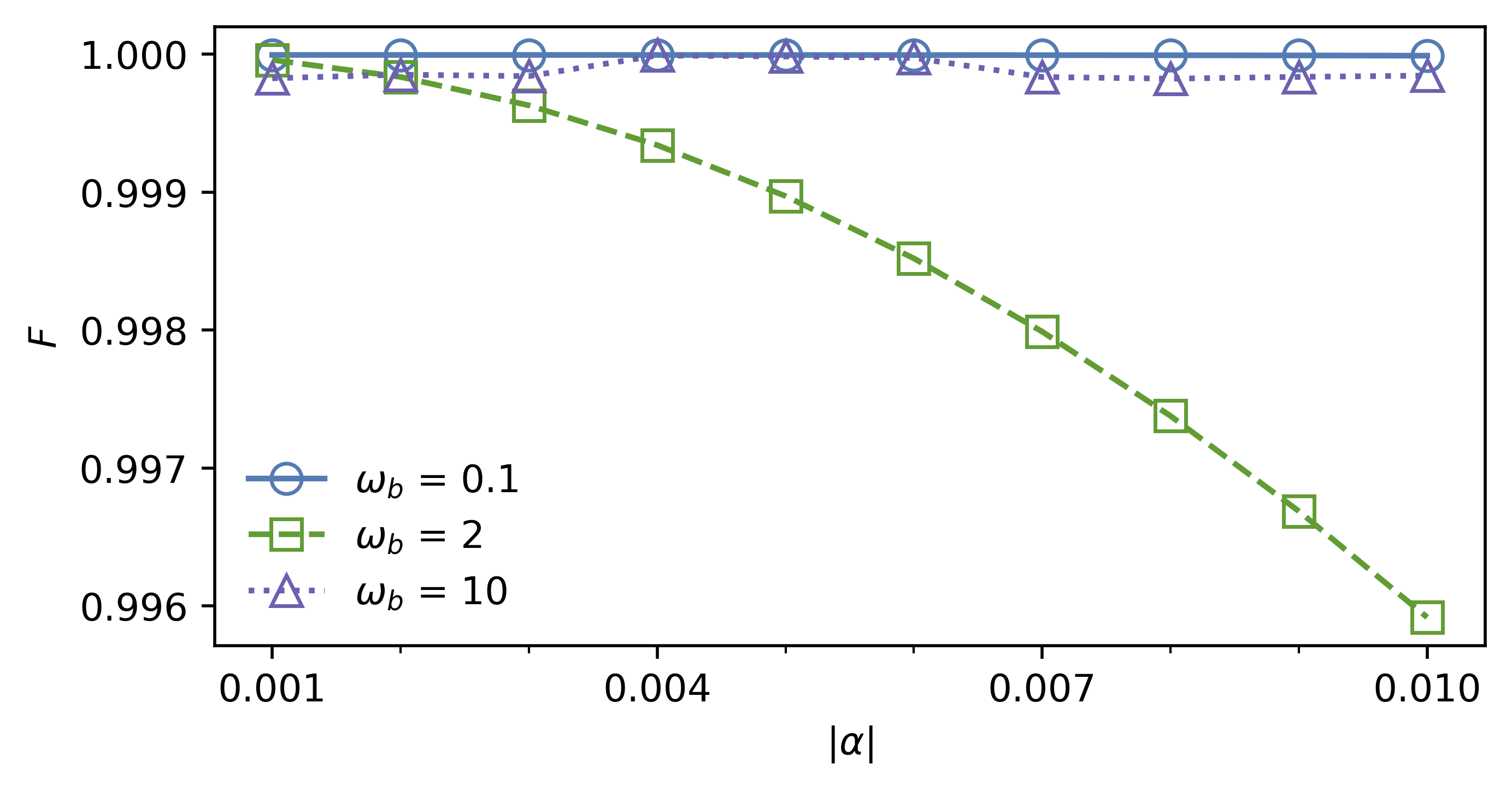}
	\caption{Fidelity $F$ vs. coupling strength $|\alpha|$ for bosonic frequencies $\omega_b = 0.1$ (sub-resonant), 2.0 (resonant), and 10.0 (super-resonant) at fixed system frequency $\omega_0 = 2$. High fidelity is maintained for small $|\alpha|$ across all cases, while the resonant case exhibits the most pronounced decay. These results delineate the parameter regime in which the dressing transformation remains accurate.}
	\label{fig2}
\end{figure}
Fig.~\ref{fig2} shows the fidelity $F$ versus coupling strength $|\alpha|$ for three representative bosonic frequencies: $\omega_b = 0.1$ (sub-resonant), $2.0$ (resonant), and $10.0$ (super-resonant), with fixed system frequency $\omega_0 = 2$. In all cases, the fidelity approaches unity for small $|\alpha|$, confirming the validity of the first-order approximation in this regime. In these simulations, the anisotropy parameter is chosen to satisfy the first-order decoupling condition, i.e., $\lambda = (\omega_0-\omega_b)/(\omega_0+\omega_b)$. This choice isolates the effect of higher-order corrections in the dressing approximation, which are responsible for the observed deviations from ideal evolution. As $|\alpha|$ increases, deviations from unity become visible due to the increasing influence of higher-order corrections. The reduction in fidelity is most pronounced near resonance ($\omega_b = \omega_0$), where system--bath interaction is strongest, and becomes much weaker in off-resonant regimes. The results delineate a broad parameter regime in which the transformation maintains high-fidelity unitary evolutions. In particular, fidelities above 0.999 are maintained in off-resonant regimes even for moderately large $|\alpha|$, demonstrating the robustness of the proposed scheme.

\begin{figure}[!tb]
	\centering
	\includegraphics[width=\columnwidth]{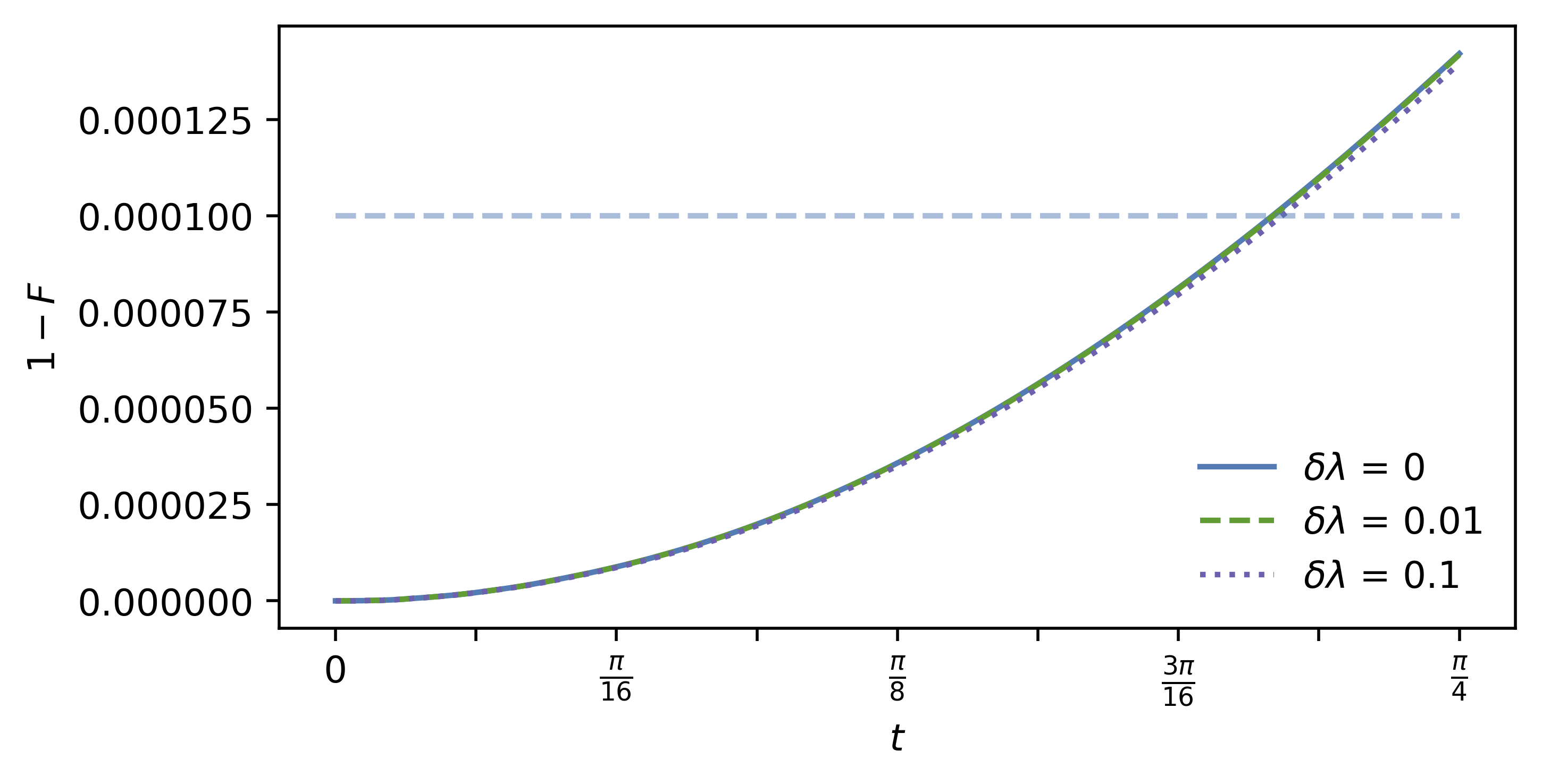}
	\caption{Infidelity $1-F$ versus evolution time $t$ for different anisotropy deviations $\delta\lambda$, at resonance $\omega_0=\omega_b=2$ and with coupling strength $|\alpha|=0.01$. The curves nearly overlap, indicating that small deviations from the ideal anisotropy have little effect on the fidelity decay and only a negligible effect on the corresponding tolerance timescale in the weak-coupling regime.}
	\label{fig3}
\end{figure}
In Fig.~\ref{fig3}, we further present the infidelity $1-F$ as a function of the evolution time $t$ for several deviations from the ideal anisotropy $\delta\lambda$, at resonance $\omega_0=\omega_b=2$ and for a weak coupling strength $|\alpha|=0.01$. As shown in the figure, the curves for different $\delta\lambda$ nearly overlap over the plotted time interval. This indicates that, in the weak-coupling regime where the first-order dressing approximation remains valid, small deviations from the ideal anisotropy have only a minor effect on the infidelity dynamics. In particular, the characteristic time at which the infidelity reaches a given tolerance $\epsilon=1\times10^{-4}$ is almost unchanged for different $\delta\lambda$. Therefore, the relevant finite time window for noise-resilient quantum computation is set primarily by the accumulation of higher-order residual interactions.

These results also help clarify the interpretation of our scheme and its analogy with one-way quantum computing mentioned above. The pre-established system--bath entanglement can be viewed as a finite resource that supports the effective cancellation of the leading system--bath coupling during the evolution. As long as the first-order dressing approximation remains valid, the dynamics follows the target noiseless evolution closely. Over longer times, higher-order residual couplings gradually accumulate and eventually lead to a noticeable loss of fidelity, thereby defining a finite time window over which noise-resilient computation can be sustained.

\section{Conclusions}
Decoherence remains a central challenge for practical quantum computing. Here we propose a distinct framework for noise-resilient quantum computing, in which pre-established system–bath entanglement is incorporated into the initial state and exploited as a resource to cancel out the dominant decoherence, in close analogy with the role of entanglement in one-way quantum computing. By appropriately initializing the system in such an entangled state, the dominant decoherence is effectively cancelled out, enabling quantum computation to proceed without additional dynamical control or error-correction overhead. We illustrate this framework using anisotropic spin–boson models, where the system dynamics is effectively decoupled from the environment within a first-order approximation, and we further confirm its validity through numerical simulations.

\section*{Acknowledgements}
This work is supported by the Natural Science Foundation of Shandong Province (Grant No. ZR2024MA046); the Fundamental Research Funds for the Central Universities (Grant No. 202364008); MCIN/AEI/10.13039/501100011033 (Grant No. PID2021-126273NB-I00); the European Regional Development Fund (``A way of making Europe"); the Basque Government (Grant No. IT1470-22); the Ministry for Digital Transformation and Civil Service of the Spanish Government through the Quantum Spain project (QUANTUM ENIA call); and the European Union through the Recovery, Transformation and Resilience Plan – NextGenerationEU, under the Digital Spain 2026 Agenda. Y. Y. Xie acknowledges additional support from China Scholarship Council (No. 202406330066).

\end{document}